\documentclass[twocolumn,showpacs,preprintnumbers,amsmath,amssymb]{revtex4}

\usepackage[dvipdfm]{graphicx}
\usepackage[dvipdfm]{color}
\usepackage{array, bm}

\begin{document}

\newcommand{\Ptorus}{P_n^{\mathrm{torus}}}
\newcommand{\PsiTorus}{\psi^{\rm torus}}
\newcommand{\PsiReg}{\psi^{\rm reg}}

\title{Recovery of 
  chaotic
  tunneling due to destruction of dynamical
 localization by external noise}

\author{Akiyuki Ishikawa}
\author{Atushi Tanaka}
\author{Akira Shudo}
\address{Department of Physics, Tokyo Metropolitan University,
  Minami-Osawa, Hachioji, Tokyo 192-0397, Japan}

\date{\today
}

\begin{abstract}
Quantum tunneling in the presence of chaos is analyzed, focusing
especially on the interplay between quantum tunneling and dynamical
localization. We observed flooding of potentially existing tunneling
amplitude by adding noise to the chaotic sea to attenuate the
destructive interference generating dynamical localization.  This
phenomenon is related to the nature of complex orbits describing
tunneling between torus and chaotic regions. The tunneling rate is
found to obey a perturbative scaling with noise intensity when the
noise intensity is sufficiently small and then saturate in a large
noise intensity regime. A relation between the tunneling rate and the
localization length of the chaotic states is also demonstrated. It is
shown that due to the competition between dynamical tunneling and
dynamical localization, the tunneling rate is not a monotonically
increasing function of Planck's constant. The above results are
obtained for a system with a sharp border between torus and chaotic
regions. The validity of the results for a system with a smoothed
border is also explained. 

\end{abstract}

\maketitle

\section{Introduction}

Since the dawn of quantum theory, the tunneling effect has been recognized as a representative of genuine quantum phenomena. This does not necessarily imply that the tunneling effect has nothing to do with underlying classical dynamics. What recently attracts much interest concerns the fact that quantum tunneling in multidimensional systems or more specifically in non-integrable systems strongly reflects the phase-space structures of the corresponding classical system%
 \cite{BTU,Creagh98,TU,Frischat95,Creagh96,
Creagh06,SI95,Takahashi00,BSU,Mouchet01,Levkov07}.

What is specific in generic nonintegrable systems which are neither completely integrable nor fully chaotic is that phase space is divided into infinitely many invariant components and they are generally intermingled with each other. 
The transition between different invariant components is forbidden in
classical dynamics, and invariant components form dynamical barriers
in phase space. On the other hand, in quantum mechanics, the
transition through dynamical barriers thus formed may be allowed as a
result of quantum effects. In particular, based on the notion of
quasimodes \cite{Arnold}, the quantum transition between congruent
tori which are symmetrically formed in phase space is called {\em
  dynamical tunneling} in the literature \cite{DH}. 

Dynamical tunneling happens in mixed-types phase space where invariant
tori and chaotic regions coexist. This fact immediately invokes a
question about the possible roles of chaos in the process of dynamical
tunneling. The notion of {\em chaos-assisted tunneling} has been
proposed to capture the situation where, besides quasi-doublet states
localized on symmetrically formed congruent tori, chaotic states take
part in the tunneling process \cite{TU}.
Semiclassical analysis in the time domain also reveals that a bunch of
tunneling trajectories associated with chaos, or more precisely the
Julia set in complex plane, is involved in dynamical tunneling in
mixed phase space \cite{SI95,SI02}. What was commonly emphasized there was an aspect that {\em chaos enhances dynamical tunneling.}

On the contrary, we here focus on another aspect of the role of chaos. In particular, it may happen that {\em dynamical tunneling is suppressed even in the presence of chaos.} As explained below, there exist two competing effects behind the dynamical tunneling process, both of which originate from the nature of complex classical trajectories describing the tunneling transition from torus to chaos. 

As shown in Refs.~\cite{SI95,SI02}, 
exponentially many complex trajectories, not a single instanton path
as in integrable tunneling, contribute to the semiclassical time
evolution operator which allows us to evaluate the tunneling amplitude
from torus to chaotic regions.  Such complex trajectories start from
an initial state placed in the torus region and reach a final state in
the chaotic sea. An important fact is that in the first stage in their
itinerary these complex orbits have non-small imaginary components and
move in the complex plane, whereas the orbits are soon attracted by
the real plane 
along stable manifolds 
and then move very close to the real plane in the later stage. The rate of the attraction is exponentially fast, reflecting the fact that the behavior of the orbits is 
controlled by the stable manifolds of unstable periodic orbits in the
real phase space \cite{SI95,SI02}. 
These observations imply that complex trajectories describing the
tunneling transport  have {\em amphibious}
character: they act as tunneling orbits when they stay in the torus region and behave as if they are almost real orbits when reaching 
the chaotic sea \cite{ITS07}.
A well recognized aspect of dynamical tunneling, that is, the
enhancement of tunneling, is explained by the fact that the number of
tunneling orbits is exponentially many \cite{SI95,SI02}.

However, their contributions to the tunneling amplitude are merely 
``potentially existing''. 
This is because 
the time evolution in the later stage may involve
opposite effects on the transport property, especially when
{\em dynamical localization} governs the quantum dynamics in the chaotic
region \cite{CCFI,Shep,FGP}.
The amphibious nature of complex orbits
predicts that the tunneling penetration through integrable barriers
and the dynamics in the chaotic sea are not independent of each other,
rather they must be closely related.
Therefore, we expect that if the spreading of wave function is
suppressed due to dynamical localization in the chaotic sea, which
appears as a result of destructive interference effects, the tunneling
transport is simultaneously suppressed.
In order words, potentially existing tunneling orbits are
exponentially many, but they interfere with each other to form
localization.
This is our working hypothesis and the purpose of this paper is just to verify that there exists strong interplay between dynamical tunneling and dynamical localization.

In order to confirm our hypothesis, that dynamical tunneling is
significantly reduced as a consequence of dynamical localization, our
strategy is to apply noise to destroy the destructive interference
generating dynamical localization. The idea is essentially the same as
the one which has been used to see that the recovery of classical
diffusion occurs when external noise is applied to dynamically
localized states \cite{OAH,ATI}. In the present case, to avoid noise-activated hopping from torus to chaos, we selectively add noise only to the chaotic region with the torus region being untouched.

We should mention that the interplay between dynamical tunneling and
dynamical localization is crucial to understand the emergence of {\em
  amphibious eigenstates} \cite{HKOS,BKM}.  In \cite{HKOS,BKM} they
suggested the competition between the effective Heisenberg time and
the tunneling time is important to determine whether eigenfunctions
show amphibious nature. The former time scale concerns dynamical
localization and the latter dynamical tunneling.
  Therefore, the {\it flooding} of eigenfunction found
  in \cite{HKOS,BKM}
  can be interpreted as a consequence of the flooding
  of potentially existing  
  tunneling orbits.

This paper is organized as follows. 
In Sec.~\ref{sec:phasespaceenginnering}, we present an area-preserving map that is designed to have phase space clearly separated into torus and chaotic regions. 
To verify our hypothesis stated above, a noise term is added. The noise term is characterized by two parameters: the strength $\epsilon$ and the distance $l$ from the border between torus and chaotic sea. In Sec.~\ref{sec:CAT}, we first investigate the case without noise as a reference for our subsequent argument about the case with noise. We especially show two typical behaviors of tunneling transition and the sensitive dependence of the tunneling amplitude on the variation in the system parameters. In Sec.~\ref{sec:expDecay}, we provide numerical evidence demonstrating that the introduction of the noise qualitatively changes the nature of tunneling. We define a tunneling decay rate $\gamma$, which will be used to quantify the rate of tunneling penetration, and then show that it is strongly enhanced if noise is imposed. In Sec.~\ref{sec:p2pi}, we examine how the noise strength affects the tunneling decay with the smallest $l$, i.e., the noise being applied to the whole chaotic sea. For small $\epsilon$, $\gamma$ increases perturbatively as $\epsilon^2$ increases. For large $\epsilon$, $\gamma$ saturates to a certain ``classical value'' and the parameter sensitivity of $\gamma$ disappears. In Sec.~\ref{sec:Ldependence}, we focus on the $l$-dependence on $\gamma$, where we show that a characteristic scale of $l$  coincides with the localization length of the chaotic sea. An interesting finding is that there exists a region of $l$ in which the tunneling rate decreases even though Planck's constant becomes large. This is exactly because of the strong interplay between dynamical tunneling and dynamical localization. In Sec.~\ref{sec:Diffraction}, we examine whether or not our result is valid for systems with a smoothed border between torus and chaotic regions.
We show that almost all results obtained in cases with clearly separated phase space hold even in such generic situations. Finally, in Sec.~\ref{sec:discussion}, we discuss our results and summarize the paper.

\section{Designing phase-space for chaotic tunneling}
\label{sec:phasespaceenginnering}

Classical phase space in a mixed system is very complex in general;
besides chaotic components, a variety of invariant components such as
Kolmogorov-Arnold-Moser (KAM) circles, islands of stability, or
cantori live together in a single phase space. Great difficulties
arise not only in classical but also in quantum dynamics in mixed
systems due to the immense complexity of phase space.  

In order to avoid complicated effects due to the mixture of different
types of invariant structures, one strategy is to make the phase space
``clean.'' Designing sharply divided phase space in a class of
piecewise linear map is in this spirit \cite{Woj}, and the recent
finding of mushroom billiards also provides a good testing ground for
a precise understanding of mixed systems \cite{Bu}. 

\subsection{Model system}
\label{sec:referenceSystem}

\begin{figure}[tb]
  \centering
  \includegraphics[width=8.5cm, clip]
  {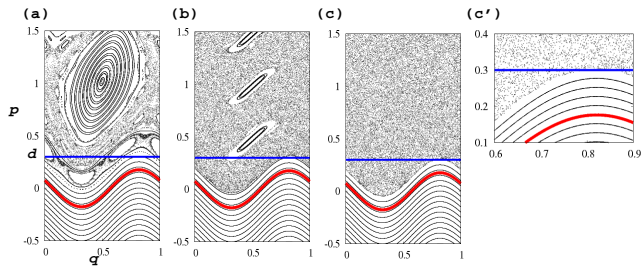}
  \caption{%
    (Color online)
    Phase space portraits of the classical map~\eqref{KickedHamiltonian}
    with $T(p)=T_0(p)$ (Eq.~\eqref{eq:defT0}).
    The parameters are set to be $d=0.3$, $\omega = 0.6418\cdots$,
    $k=2$ and (a) $s=0.5$, (b) $s=2$, and (c) $s=4$. 
    The smoothing parameter $\beta=100$ is chosen so that the border between torus and chaotic regions is clear. Note that the border is not sharp in the strict sense (see text). 
    A magnification of the area around the border in the case of $s=4$ is shown in (c').  For small $s$, there are big islands in the chaotic sea and also around the border. On the other hand, if $s$ is sufficiently large, e.g., $s=4$, these islands become so small that the influence of these almost invisible islands on dynamical tunneling is negligible.
  }
  \label{FIG:PhaseSpaces_s}
\end{figure}

We here present our system whose classical phase space is clearly separated into torus and chaotic regions. Let us consider the one-dimensional kicked rotor described by the Hamiltonian
 \begin{equation}
   \label{KickedHamiltonian}
  H(p, q, t) = T(p) + V(q)\sum _{n=-\infty} ^{\infty}\delta (t-n),
 \end{equation}
 where $p$ and $q$ are the momentum and the position (angle) of rotor, respectively.
 The corresponding classical dynamics is reduced to the area preserving map
 \begin{equation}
   \label{BasicMap}
  F :
   \left(
    \begin{array}{c}
     p_{n+1}\\
     q_{n+1} 
    \end{array}
      \right) = \left(
      \begin{array}{c}
        p_n - V'(q_n)\\
        q_n + T'(p_{n+1})
      \end{array}
    \right).
 \end{equation}

To realize the phase space which is clearly separated into regular and chaotic regions, we take the kinetic and potential terms in the following way: 
for the kinetic term, we have
 \begin{align}
   \label{eq:defT0}
   T_0(p)&
   = \frac{1}{2} s(p-d)^2\theta_{\beta}(p-d) + \omega (p-d),
 \end{align}
 where $\theta_{\beta}(x)$ denotes a smoothed step function
 with a parameter $\beta(>0)$: 
 \begin{align}
   \label{eq:defTheta}
   \theta_{\beta}(x)
   \equiv 
   \frac{1}{2}\left\{1 + \tanh (\beta x)\right\},
 \end{align}
 e.g., $\theta_{\beta}(x)\to 0$ as $x\to-\infty$ and 
 $\theta_{\beta}(x)\to 1$ as $x\to+\infty$.
  $\lim_{\beta \to \infty}\theta_{\beta} (x)$ gives the Heaviside step function.
 For $p\ll d$, $T_0(p)\simeq\omega (p -d)$ is a kinetic term for 
 a linear rotor with a rotation number $\omega$.
 On the other hand, above the ``border'' $p=d$ in the momentum space,
 the quadratic term  $s (p-d)^2/2$ appears in the kinetic term.
 Throughout this paper we employ the following potential term
 \begin{align}
   \label{eq:defV}
   V(q) &= \frac{k}{4\pi^2} \cos(2\pi q).
 \end{align}
 We therefore have, for $p\ll d$, 
 \begin{equation}
   \label{eq:IntegHamiltonian}
   H\approx \omega (p-d) + \frac{k}{4\pi^2}\cos(2\pi q)\sum_{n=-\infty}^{\infty} \delta (t-n)
 \end{equation}
 which is completely integrable, 
 and, for $p\gg d$, 
 \begin{equation}
   H\approx
   \frac{1}{2}s(p-d)^2 + \omega (p-d)
   + \frac{k}{4\pi^2}\cos(2\pi q)\sum_{n=-\infty}^{\infty}  \delta (t-n),
 \end{equation}
 which is reduced to the standard mapping with a nonlinearity
 parameter $s k$  under an appropriate scaling of the Hamiltonian and
 a shift in the momentum space.  Figure~\ref{FIG:PhaseSpaces_s} shows
 examples of the classical phase space with several different values
 of $s$.In the following, we will examine the case of $s=4$ where
 phase space is clearly separated into torus and chaotic regions. As
 shown in Fig.~\ref{FIG:PhaseSpaces_s} (c), when $s=4$ the islands
 around the border between torus and the chaotic regions are almost
 invisible.  However, it should be noted that the border is not as
 sharp as in a previously studied class of piecewise linear map
 \cite{Woj} or mushroom billiards \cite{Bu}. One can find tiny island chains in much more magnified figures.

 \subsection{Introduction of the effect of noise}
 \label{sec:IntroNoise}

As is well known, dynamical localization occurs as a result of
destructive interference in the chaotic sea
\cite{CCFI,Shep,FGP}. Conversely, once quantum interference is
destroyed in some way, classical diffusion is recovered \cite{OAH,ATI}. In a similar way, we will verify that chaotic tunneling is suppressed by dynamical localization in the chaotic sea by examining whether or not potentially existing chaotic tunneling is recovered if destructive interference creating dynamical localization is attenuated.

A simple way to destroy quantum interference is, as done in numerical
tests to see the recovery of classical diffusion \cite{OAH,ATI}, to apply noise to chaotic seas. However, the introduction of noise needs to be done carefully and should only be limited to chaotic regions, otherwise the tunneling process will be contaminated by the thermal hopping between the torus and the chaotic seas. In order to avoid the transition from the torus region into chaotic sea, noise will be applied only to the region that is far from the torus, say $p > L$ ($>d$). We employ the following modification to the kinetic term to realize such momentum-dependent noise:
 \begin{equation}
   \label{eq:noise}
   T_\epsilon (p,n) 
   = T_0(p) + \epsilon\xi_n(p-L)\theta_{\beta}(p-L) 
  ,
 \end{equation}
 where $\epsilon$ is the noise strength and $\xi_n$ is a stochastic variable.
 We assume that $\xi_n$ obeys a Gaussian distribution:
 \begin{equation}
  \langle \xi_n \rangle = 0, \quad  \langle \xi_n \xi_m \rangle = \delta_{mn},
 \end{equation}
 where $\langle\cdot\rangle$ denotes the
   ensemble average.

 \subsection{Quantum map}

 The quantum dynamics associated with the classical map Eq.~(\ref{BasicMap}) is presented following the standard quantization procedure of quantum maps.  Due to the periodicity of the map Eq.~(\ref{BasicMap}) in $q$, the periodic boundary condition (with a period $1$) is imposed on the wave function in the position representation. We also impose the periodic boundary condition in the momentum space $-\frac{W}{2}\le p \le \frac{W}{2}$.
 As a result of these boundary conditions, the effective Planck's constant is thus quantized as \begin{equation}
   h = \frac{W}{N},
 \end{equation}
 where $N$ is the dimension of the Hilbert space of the system. Furthermore, in order to avoid the finite size effect in the diffusion process, an absorbing boundary is set at $p =  p_{\rm cutoff}$. We deliberately choose $p_{\rm cutoff}$ large enough to ensure the convergence of the characteristics of dynamical tunneling
 from torus to chaotic sea \cite{EN1}.
 A unit time evolution of the state vector $|\psi_n\rangle$ is
 described by the quantum map
\begin{equation}
  \label{eq:qmap}
  |\psi _{n+1} \rangle = \hat{U}|\psi_n\rangle,
\end{equation}
where $\hat{U}$ is the time evolution operator for a unit time interval
 \begin{equation}
   \Hat{U} = \exp \left(-\frac{2\pi i}{h} T(\Hat{p}) \right)
   \exp \left(-\frac{2\pi i}{h} V(\Hat{q}) \right).
 \end{equation}

\section{%
 Tunneling doublet,
 suppression of decay, 
 and dynamical localization in the system without noise
 }
\label{sec:CAT}

 \begin{figure}[tb]
   \centering
   \includegraphics[width=8.0cm, clip]
   {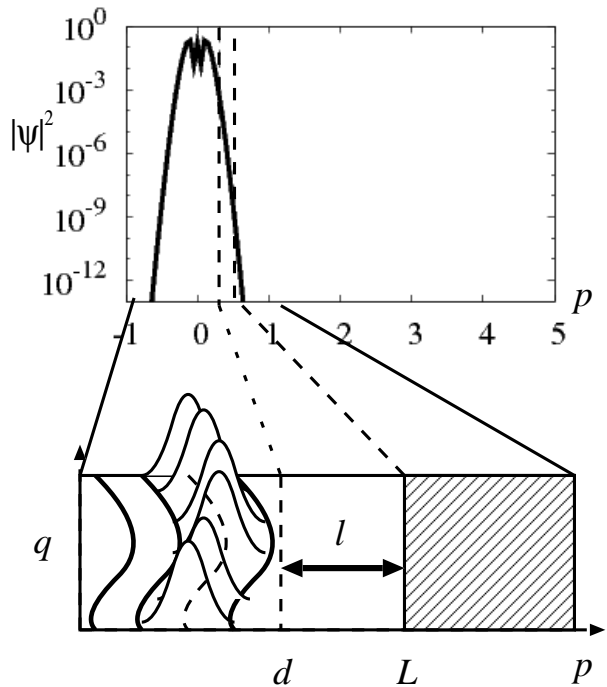}
   \caption{%
     Probability distribution of initial
     wave function in the momentum space (upper) and its schematic illustration 
     in phase space (lower). We impose noise on the shaded domain.}
   \label{FIG:Schematic}
\end{figure}

 \begin{figure}[tb]
   \centering
   \includegraphics[width=8.5cm, clip]
   {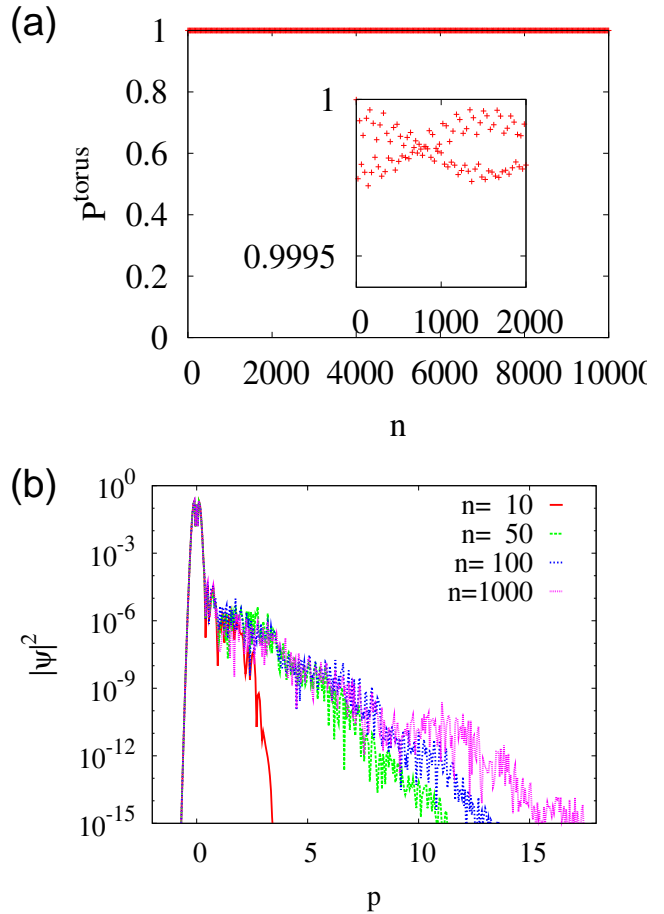}
   \caption{%
     (Color online)
     An example of the off-resonant case
     ($s = 4.2566\cdots$) with $h=1/20$, $W=40$ and $p_{\rm cutoff} = 18$. 
     Other parameters are the same as Fig.~\ref{FIG:PhaseSpaces_s}. At
     $n=0$ almost all of the probability is localized in the torus
     region. 
     (a) $\Ptorus$ as a function of time step $n$. During the time
     evolution, $\Ptorus$ keeps almost unity value. In the inset, we
     magnify the tiny oscillation of $\Ptorus$. 
     (b) Snapshots of the momentum distribution $|\psi_n(p)|^2$. The
     border between the torus and chaotic regions is at $d=0.3$.
       The diffusion in the chaotic sea is suppressed by dynamical
       localization, so the probability distribution tends to a
       stationary distribution with an exponentially decaying tail,
       which is characteristic of dynamical localization in the
       chaotic sea.
   }
  \label{FIG:TimeEvo-NonReso}
\end{figure}

\begin{figure}[tb]
  \centering
   \includegraphics[width=8.5cm, clip]
   {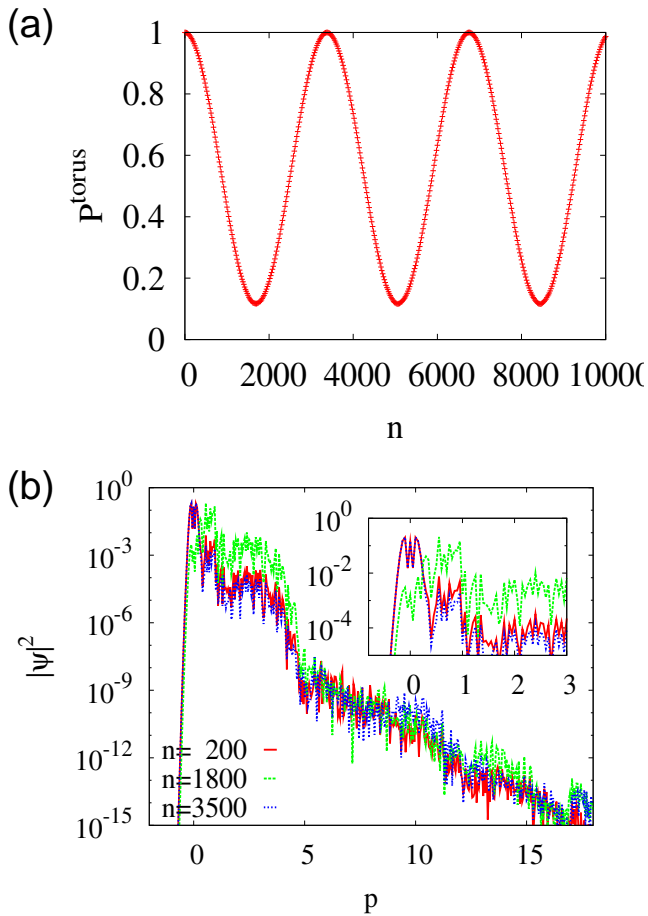}
  \caption{%
    (Color online)
    An example of the resonant case. Except for $s$( $= 4.2178\cdots$),
    the values of the parameters are the same as 
    in Fig.~\ref{FIG:TimeEvo-NonReso}.
    (a) $\Ptorus$ as a function of time step $n$. 
    A large oscillation ranging from 0.1 to 1.0 is observed. 
    (b) Snapshots of the momentum distribution $|\psi_n(p)|^2$
    at $n=200, 1800$ and $3500$, each of which 
    corresponds to an extremum in the curve shown in (a).
    Dynamical localization occurs in the chaotic
    region, as it does for the off-resonant case.
    In the inset, tunneling oscillation between the torus region
    $p<d$ and a part of the chaotic sea $d<p<5$ are magnified.
  }
  \label{FIG:TimeEvo-Reso}
\end{figure}
 \begin{figure}[tb]
  \centering
  \includegraphics[width=8.5cm, clip]
   {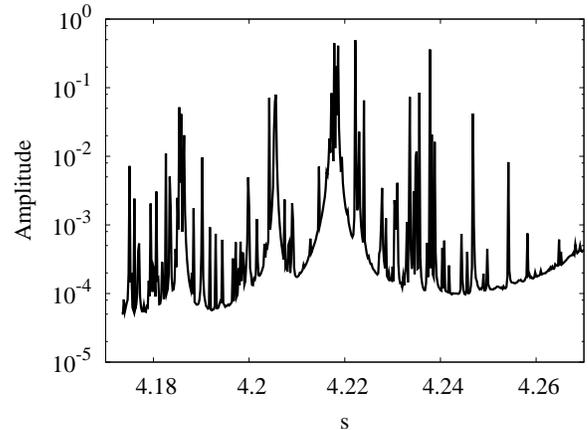}
  \caption{The oscillation amplitude of $\Ptorus$ as a function of $s$.}
  \label{FIG:amp-s}
\end{figure}

 \begin{figure}[tb]
  \centering
  \includegraphics[width=8.5cm,clip]
   {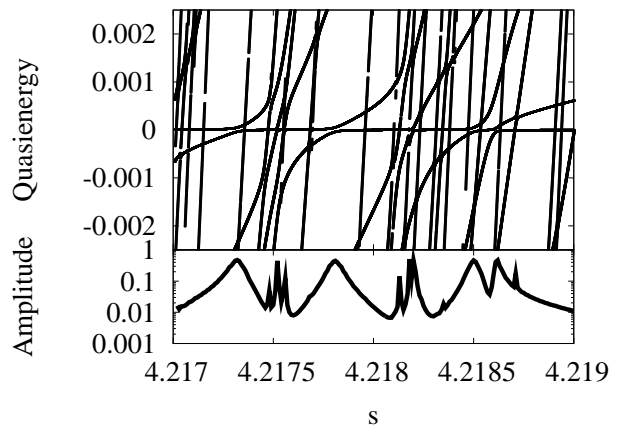}
  \caption{Quasienergies of torus and chaotic states (upper) and
    the magnification of oscillation amplitude of $\Ptorus$ in
    Fig.~\ref{FIG:amp-s}
    as a function $s$ (lower).
    Only the quasienergies of the eigenstates whose momentum
    expectation values fall within the range $-0.2<p<6.0$ are
    shown \cite{EN4},
    because the chaotic states which have large contribution 
    to the oscillation are expected to be near the torus state.
    A line running in the horizontal direction represents an
    eigenenergy associated with a torus eigenstate. It forms many
    avoided crossings with energy levels corresponding to chaotic
    states. Note that peaks of oscillation amplitude appear when
    avoided crossings occur.  
    }
  \label{FIG:s-eigen_energy}
\end{figure}
Before studying the case where the noise term is present, we show that the tunneling process from the torus to the chaotic region is sensitive to the change in system parameters, even though the classical phase space of our system has carefully been designed to remove resonance islands. The result tells us that the quantum interference effect is very relevant in the issue of dynamical tunneling, and it strongly affects the transition process 

Hereafter, the time evolution of tunneling is quantitatively monitored using
\begin{equation}
 \Ptorus = \int^{p_b} _{p_a} \left| \psi_n(p) \right|^2 dp,
\end{equation}
where $\psi_n (p)$ is the wave function in the momentum representation
at the $n$-th step, and parameters $p_a$ and $p_b$ are appropriately
chosen to enclose the torus region. We use $p_a = -W/2$  and  $p_b =
d$ in the following \cite{EN2}.

Throughout this paper, the initial wave function is taken as an
eigenstate of the linear rotor  Eq.~\eqref{eq:IntegHamiltonian}. As
explained in Appendix~\ref{sec:QuantizeTorus}, the initial
wave function $\PsiTorus{}(q; p_{\rm c})$ in the position
representation is specified by $p_{\rm c}$, where  $p_{\rm c}$
represents the ``average'' of the momentum of the torus \cite{EN3}.
The initial state is placed just below the border between the torus and the chaotic sea, i.e. $p_{\rm c}=0$  (see Fig.~\ref{FIG:Schematic}). This is because the tunneling amplitude becomes too small compared to the numerical precision if the initial state is placed too far from the border.   

As mentioned above, even with a small change in parameters, $\Ptorus$ changes rather drastically and there appears a variety of oscillatory patterns in the plot of $\Ptorus$. In the following, we show two typical examples. As seen below, in spite of only slightly different values of $s$, oscillatory patterns are entirely different.  Note that variation in $s$ does not have any effect on the integrable region. It only affects chaotic regions. The classical phase space portrait is therefore almost unchanged under a small change in $s$. 

First we depict the time evolution of $\Ptorus$ in the case of $s =
4.2566\cdots$ (see Fig.~\ref{FIG:TimeEvo-NonReso}). We hereafter call
this the {\em off-resonant} case~\cite{endnote:resonance}. 
$\Ptorus$ keeps almost unity value
as shown in Fig.~\ref{FIG:TimeEvo-NonReso}(a), which means that most
of the probability amplitude is confined within the torus region. On
the other hand, as shown in Fig.~\ref{FIG:TimeEvo-NonReso} (b), the
snapshots of the log-scaled probability distribution in the momentum
representation, once the tunneling transition from the torus to
chaotic region occurs, the wave function starts to diffuse into the
chaotic sea. However, the diffusion is gradually saturated and the
wave function asymptotically approaches a stationary shape. This is a
result of dynamical localization. In other words, almost all of the
probability amplitude that tunnels out of the torus region is confined
in a region $d < p < d + \xi$. Here $\xi$ represents the localization
length in the chaotic sea \cite{Shep}.

Next, we slightly change the parameter to $s = 4.2178\cdots$ (see
Fig.~\ref{FIG:TimeEvo-Reso}). In contrast with the previous case shown
in Fig.~\ref{FIG:TimeEvo-NonReso}(a), $\Ptorus$ exhibits a periodic
oscillation (see Fig.~\ref{FIG:TimeEvo-Reso}(a)).  In accordance with
the oscillation of $\Ptorus$, as shown in
Fig.~\ref{FIG:TimeEvo-Reso}(b), the momentum distribution also
exhibits a recurrent pattern. Initially, the wave function is localized
within the torus region. As time elapses, most of the probability
amplitude escapes to the chaotic region. After that, the wave function
comes back to the torus region. Such a recurrent oscillation strongly
suggests the emergence of a tunneling doublet composed by a torus and
a chaotic state. We call this type of behavior the {\em resonant
case}~\cite{endnote:resonance}.

To see large fluctuation with the variation in $s$ more explicitly, we
present in Fig.~\ref{FIG:amp-s} the amplitude of the time-dependent
oscillation of $\Ptorus$  as a function of $s$. Such an erratic
oscillation pattern reminds us of chaos-assisted tunneling, where the
tunnel splitting between symmetry-related torus states exhibits strong
fluctuation with variation in parameters \cite{BTU,TU}. In the case of chaos-assisted tunneling, the fluctuation is induced by avoided crossings that involve the torus and the chaotic states.

Now we explain the origin of the sensitivity of the tunneling to the
variation in $s$ in the present case. This is also due to the
sensitivity of accidental degeneracies between torus and chaotic
states.
As evidence, we compare the quasienergies and the
oscillation amplitude of $\Ptorus$ in
Fig.~\ref{FIG:s-eigen_energy}. 
Since $s$ mainly affects the dynamics
in the chaotic sea,
the quasienergies of the torus states keep almost constant and form
avoided crossings with many energy levels corresponding to chaotic
states.
Around these avoided crossings, coherent oscillation between torus and
chaotic states occurs. In Fig.~\ref{FIG:s-eigen_energy},
we clearly
see the correspondence between peaks of the oscillation amplitude and
positions of avoided crossings.
Note that the period of the tunneling oscillation is
well described by the formula $T_{\rm osc} = 2\pi\hbar/\Delta$, where
$\Delta$ is the gap of quasienergies.

We remark on the influence of the absorbing boundary. The distance
between the positions of the border $d$ and the absorbing boundary
$p_{\rm cutoff}$ is much larger than the localization length, 
{\it e.g.}, $p_{\rm cutoff} - d \gg \xi$. 
The probability
amplitude which reaches the absorbing boundary is exponentially
small, even if almost all of the probability amplitude
temporarily gets out of the torus region in the resonant case (see
Fig.~\ref{FIG:TimeEvo-Reso}(b)).
Hence the influence from the absorbing boundary is virtually
negligible. In this sense, the wave function is approximately bounded 
around the torus region or the nearby chaotic sea.

We summarize our findings in this section. 
In the absence of noise, the behavior of the wave function 
launched from the torus region sensitively depends on the parameter $s$ 
while classical dynamics does not. 
This is explained by the presence of the near-degeneracies of quasienergies.
In particular, we discovered a new kind of tunneling doublets that couples 
a torus state with a chaotic state that exhibits dynamical localization.

\section{
    Exponential decay
  due to 
  the breakdown of dynamical localization}
\label{sec:expDecay}

 \begin{figure}[tb]
  \centering
  \includegraphics[width=8.5cm, clip]
   {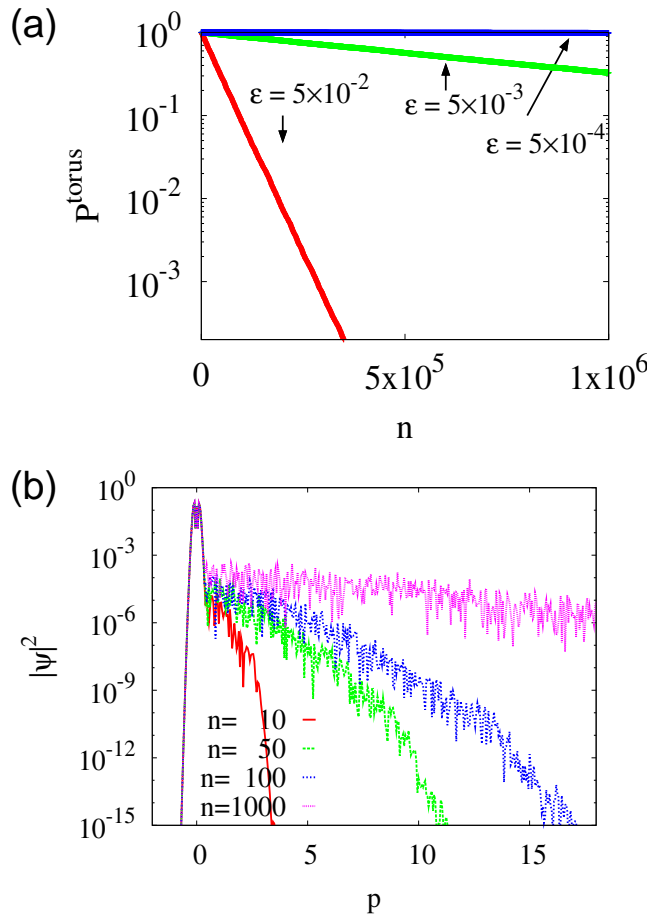}
  \caption{(Color online)
    The off-resonant case under the presence of noise
    with $L=d+0.2$. The other parameters are the same as in
    Fig.~\ref{FIG:TimeEvo-NonReso}.
    (a) $\Ptorus$ exhibits exponential decay, where
    the exponent increases as $\epsilon$ increases.
    (b) We depict snapshots of the probability distribution at
    $\epsilon = 0.005$.
    For large $n(\ge 100)$, the effect of noise becomes significant
    (cf. Fig.~\ref{FIG:TimeEvo-NonReso}(b)).
  }
  \label{FIG:TimeEvo-NonResoNoise}
\end{figure}
 \begin{figure}[tb]
  \centering
  \includegraphics[width=8.5cm, clip]
   {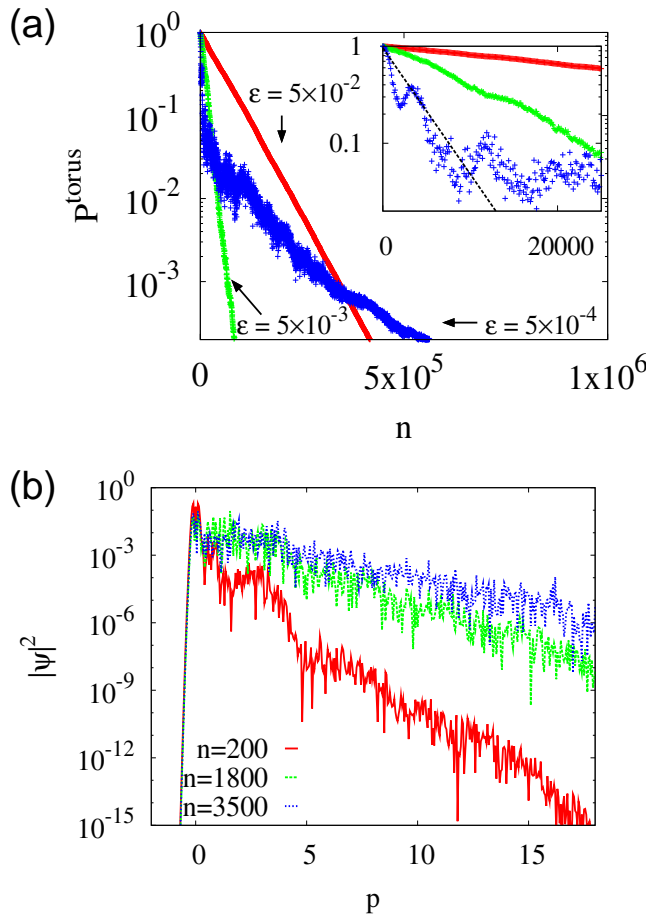}
  \caption{(Color online)
    The resonant case with noise ($L=d+0.2$). The other parameters are the same as in Fig.~\ref{FIG:TimeEvo-NonResoNoise}. 
    (a) $\Ptorus$ exhibits oscillatory decay for $\epsilon = 5\times 10^{-4}$.     To introduce a decay rate, we fit an exponential curve to the oscillatory decay for short time steps (e.g. $n\le 10^{4}$). A dashed line in the inset depicts the result of the fitting to the oscillatory decay. As the effect of noise increases, the resonant oscillation disappears and $\Ptorus$ shows exponential decay for longer time steps, where the decay rate slows down in contrast with the off-resonant case. The decay rate for the strongest noise (e.g. $\epsilon = 0.005$) is almost the same as that for the off-resonant case. (see Fig.~\ref{FIG:TimeEvo-NonResoNoise}(a)).
    (b) Snapshots of the time evolution at $\epsilon=5\times10^{-4}$  are presented.  Each time step corresponds to the time step in Fig.~\ref{FIG:TimeEvo-Reso}(b).
  }
  \label{FIG:TimeEvo-ResoNoise}
\end{figure}

As shown in the previous section, the tail of the wave function, which is initially localized and exponentially decaying towards the chaotic sea, gradually spreads over the chaotic sea as time proceeds and dynamical localization controls the nature of the wave function there. In other words, dynamical tunneling necessarily accompanies dynamical localization in the chaotic sea. 

To examine the influence of dynamical localization on the tunneling
process further, we will add external noise to the chaotic region to
attenuate dynamical localization. As is explained in
Section~\ref{sec:IntroNoise},  
the kinetic term of our kicked Hamiltonian is replaced with $T_\epsilon (p,n) $ (Eq.~\eqref{eq:noise}), where noise is applied only to the region $p > L$. To avoid contaminating the tunneling process with ``thermal agitation'', we impose the condition $L > d$. Furthermore, in this section, we make $L$ as small as possible to observe the case where dynamical localization is attenuated in the whole chaotic sea. In the following, we examine the effect of noise for both resonant and off-resonant cases.

First, for the off-resonant case ($s = 4.2566\cdots$), we can
explicitly see from Fig.~\ref{FIG:TimeEvo-NonResoNoise}(a) that
unlimited diffusion does take place \cite{EN5},
which is in sharp contrast to Fig.~\ref{FIG:TimeEvo-NonReso}. 
More importantly $\Ptorus$ decays exponentially, and its decay rate
depends on the intensity of the noise. As seen in
Fig.~\ref{FIG:TimeEvo-NonResoNoise}(a), the decay rate gets larger
with the increase in the noise intensity. The origin of the
exponential decay of $\Ptorus$ can be attributed to the tunneling
effect, since we apply noise so that it does not affect the classical
dynamics in the torus region. In other words, the exponential decay
has nothing to do with thermal hopping from the torus into the chaotic
regions. Typical snapshots of the probability distribution are shown
in Fig.~\ref{FIG:TimeEvo-NonResoNoise}(b), where the effect of noise
becomes significant at larger $n$.

Thus we show for the first time that chaotic tunneling from a
torus into chaotic regions is enhanced drastically even though the
noise is added only to the chaotic sea.  This means that the
recovery of diffusion in the chaotic region or the disappearance of
dynamical localization gives rise to drastic enhancement of dynamical
tunneling, or alternatively stated, dynamical localization suppresses
dynamical tunneling.

\begin{figure}
  \centering
  \includegraphics[width=8.5cm, clip]
   {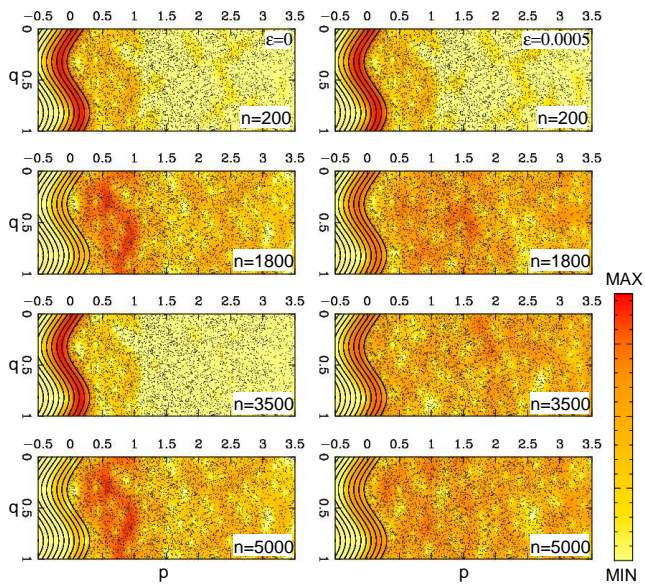}
  \caption{%
      (Color online)
      Time evolution of Husimi functions (in a logarithmic scale) in 
      the resonant case
      for $\epsilon=0$ (left column) and $\epsilon=0.0005$ (right column)
      at $n = 200, 1800, 3500$ and $5000$ (from top to bottom).
      The corresponding phase space portrait is superposed.
      The other parameters are the same as in 
      Figs.~\ref{FIG:TimeEvo-Reso} and~\ref{FIG:TimeEvo-ResoNoise}.
      For $\epsilon=0$, there occurs coherent tunneling between
      the torus region and the chaotic sea. At $n=1800$ and $n=5000$,
      the system completely tunnels out to a localized region
      in the chaotic sea. The system periodically comes back
      to the torus region (e.g., $n=3500$).
      Hence the ``pumping out'' and ``pumping back'' of wavepacket 
      between the torus region and chaotic seas
      cancels out in the absence of noise in the long time average.
      The pumping still remains for $\epsilon=0.0005$, though 
      the cancellation is broken. 
      On a short time scale, the effect of noise is negligible ($n=200$).
      At $n=1800$, the noise hinders the tunneling from the torus 
      partially, and at the same time, promotes the diffusion 
      at the chaotic sea. 
      The discrepancy due to the noise is more evident at $n=3500$.
      Although $P^{\mathrm{torus}}_{n=3500}$
      is far larger than $P^{\mathrm{torus}}_{n=1800}$
      (see the inset of Fig.~\ref{FIG:TimeEvo-ResoNoise} (a)),
      the recurrence is not quite perfect.
      In the chaotic sea, there remains a diffusive component,
      which is produced by the tunneling in the previous steps
      and fails to back to the torus region.
      This explains why 
      the pumping that occurs in the resonant case promotes the
      tunneling decay (see, Figs.~\ref{FIG:TimeEvo-NonResoNoise} (a)
      and \ref{FIG:TimeEvo-ResoNoise} (a)).
  }
  \label{FIG:pump}
\end{figure}

Second, we examine the resonant case ($s = 4.2178\cdots$), where
$\Ptorus$ periodically oscillates in time. In the presence of noise,
the behavior of $\Ptorus$ is sensitive to $\epsilon$ (see
Fig.~\ref{FIG:TimeEvo-ResoNoise}(a)).  $\Ptorus$ decays much faster
than in the off-resonant case (see
Fig.~\ref{FIG:TimeEvo-NonResoNoise}(a)). This is because dynamical
tunneling, whose restoration is due to the presence of noise, as in
the off-resonant case, is pumped by the resonance, which remains
because the noise intensity is still small enough. Snapshots of
probability distribution
are shown in Fig.~\ref{FIG:TimeEvo-ResoNoise}(b).
We also show a magnified view of Husimi functions in Fig.~\ref{FIG:pump}
to explain the pumping.
For larger $\epsilon$, $\Ptorus$
exhibits exponential decay. Furthermore, as $\epsilon$ increases, the
decay of $\Ptorus$ becomes slower, in contrast with the off-resonant
case. This suggests that the effect of resonance becomes weaker as the
noise intensity becomes larger.

These observations motivate us to introduce a decay rate $\gamma$ for $\Ptorus$  to characterize the restoration of dynamical tunneling, at least in an early stage, e.g., $n \lesssim 10^4\sim10^6$ steps in Figs.~\ref{FIG:TimeEvo-NonResoNoise} and~\ref{FIG:TimeEvo-ResoNoise}. We should remark, however, that the fitting is rather subtle for weak $\epsilon$ in the resonant case, where the ``early stage'' becomes rather short, e.g., $n \lesssim 10^4$ in the example shown in Fig.~\ref{FIG:TimeEvo-ResoNoise}.

\section{
  Noise strength dependence of chaotic tunneling
}
\label{sec:p2pi}

\begin{figure}[tb]
  \includegraphics[width=8.5cm, clip]
   {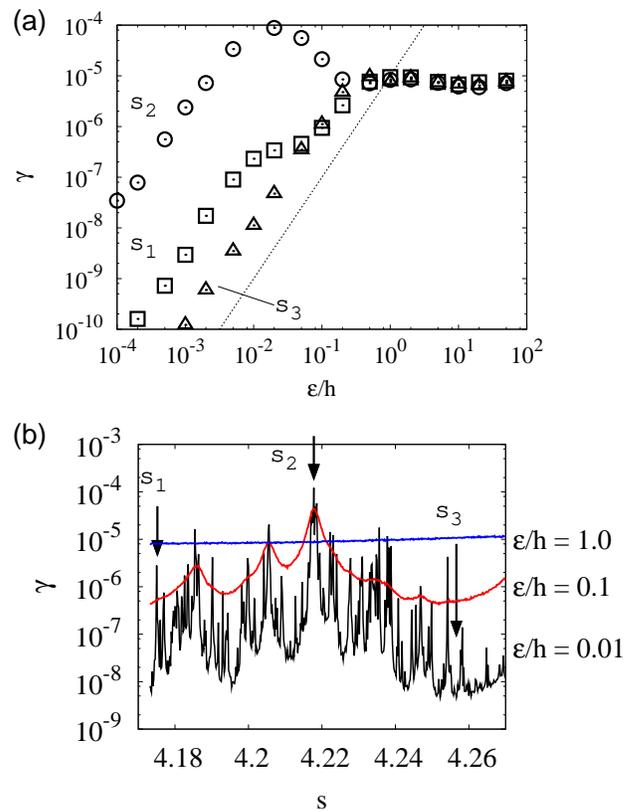}
 \caption{ (Color online)
   (a) The decay rate $\gamma$ as 
   a function of $\epsilon/h$ for different
   values of the parameter $s = s_1( = 4.175\cdots), s_2(= 4.2178\cdots)$
   and $s_3(= 4.2566\cdots)$. 
   The former two correspond to resonant cases and the latter to an off-resonant case. The other parameters are the same as in Fig.~\ref{FIG:TimeEvo-NonReso}.  When the effect of noise is weak, the tunneling decay rate obeys a perturbative behavior $\gamma \sim \epsilon^2$ (dotted line). $\gamma$ is saturated in the strong intensity regime ($\epsilon/h > 1$). (b) The decay rate $\gamma$ as a function of $s$ for different $\epsilon/h$. Arrows indicate $s_j$'s. The locations of the peaks for small $\epsilon$ coincide with the resonances of the unperturbed system (see also Fig.~\ref{FIG:amp-s}).
 } 
 \label{FIG:Gamma_s_Epsilon}
\end{figure}

In this section, we examine the dependence of the tunneling decay rate
$\gamma$, introduced in the previous section, on the noise strength
$\epsilon$. We first recall the studies on the effect of noise on
dynamical localization \cite{OAH,ATI}, in which it was reported that the diffusion rate of quantum systems when noise is imposed depends on the noise strength. More precisely, if noise is weak enough, the behavior of the diffusion rate as a function of the noise strength is understood within a perturbation theory: the diffusion rate scales with the square of noise strength. In contrast, if noise is stronger than a certain threshold, the diffusion rate saturates to a classical value. We here expect that a similar dependence of tunneling on the noise intensity would be observed. Figure~\ref{FIG:Gamma_s_Epsilon} actually verifies this prediction. Below we will explain how $\gamma$ depends on $\epsilon$ in more detail.

When noise is weak ($\epsilon/h \ll 1$), the decay rate $\gamma$ is proportional to $(\epsilon/h)^2$. This behavior can be, although not explicitly shown here, reproduced by performing a perturbation expansion with respect to a small parameter $\epsilon$. Qualitatively, exponentially decaying behavior is observed as a result of destruction of dynamical localization, therefore, as the degree of destruction is increased, the decay rate $\gamma$ grows. Furthermore, $\gamma$ strongly depends on the parameter $s$, and in particular, the locations of the peaks for finite $\epsilon$  coincide with the resonances of the unperturbed system ($\epsilon=0$), which is shown in Fig.~\ref{FIG:Gamma_s_Epsilon} (b). This means that the effect of resonance persists in the perturbative regime.

When $\epsilon$ lies in an intermediate region, the decay rate
$\gamma$ is still influenced by the effect of noise. In the
off-resonant case, $\gamma$ obeys the perturbative behavior over a
rather wide range of $\epsilon$. On the other hand, in the resonant
case, the perturbative argument is not applicable in the intermediate
region, and the rate of growth of $\gamma$ with $\epsilon$
depends on $s$. Two typical examples are shown in Fig.~\ref{FIG:Gamma_s_Epsilon}~(a)
(
 $10^{-2}< \epsilon/h < 1$
 ).
With $s=s_1$, the rate of growth of $\gamma$ with $\epsilon^2$ temporarily decreases and $\gamma$ converges to the value for the off-resonant case ($s=s_3$) as $\epsilon$ is increased. In another case, $s=s_2$ in Fig.~\ref{FIG:Gamma_s_Epsilon}~(a),  $\gamma$ temporarily decreases even with increase in $\epsilon$. A possible explanation is that the effect of resonance is attenuated by moderate noise.

We summarize the results for weak and intermediate noise intensity regions: an overall trend is that $\gamma$ increases on average, whereas the fluctuation of $\gamma$ decreases with increase in $\epsilon$. This suggests that, in the weak and intermediate regions, there exists a competition of two different interference effects; one is a constructive interference induced by resonances, and the other
a destructive interference associated with dynamical localization. The former and the latter correspond to the changes of the fluctuation and the average of $\gamma$, respectively. 

For large $\epsilon$, it is natural to expect that these interference effects 
become totally incoherent. Indeed, we numerically find that $\gamma$
is almost independent of $s$, which is shown in
Fig~\ref{FIG:Gamma_s_Epsilon}. 
What is most important is that $\gamma$ seems to converge to a certain  
value.
The existence of the plateau regime (in the noise strength dependence  
of the tunneling rate)
strongly suggests a ``classical value'' of the tunneling  
rate $\gamma$ in the incoherent limit.
If this is the case, it must be evaluated based on complex classical  
orbits.

\section{Partial destruction of dynamical localization}
\label{sec:Ldependence}
 \begin{figure}[tb]
  \centering
  \includegraphics[width=8.5cm, clip]
   {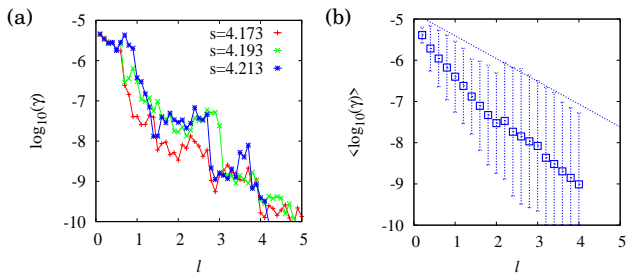}
  \caption{%
    (Color online)
    (a) The logarithm of the tunneling rate $\gamma$ as a function of $l = L-d$     for several values of $s$.
    (b) The average and variance of $\log_{10}\gamma$. 
    A theoretical line
    $\gamma_{\text{theory}}\sim e^{-2 l/\xi}$ (see text) is drawn 
    as a dashed line. 
    The parameters are $h = 1/20$ and $\epsilon/h = 5$.
  }
  \label{FIG:Gamma-L}
\end{figure}
\begin{figure}[tb]
  \centering
  \includegraphics[width=8.5cm, clip]
   {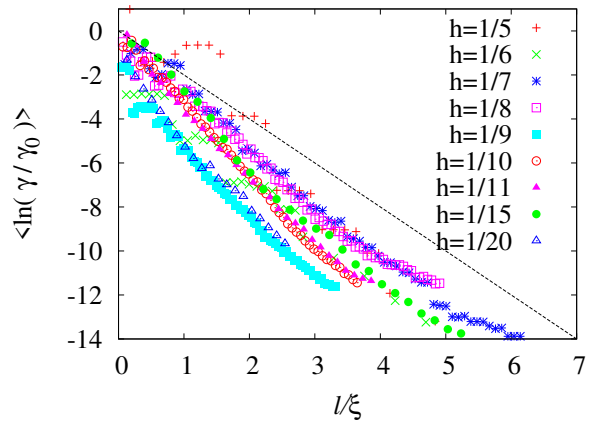}
  \caption{(Color online) The scaled exponent of the decay rate $\langle
    \log (\gamma/\gamma_0) \rangle_s$ as a function of the scaled
    width $l/\xi$.
    The dotted straight line is the relation \eqref{eq:AverageLnGamma}
   with $\alpha = 2$.
}
  \label{FIG:scaled-gamma}
\end{figure}

\begin{figure}[tb]
  \centering
  \includegraphics[width=8.5cm, clip]
   {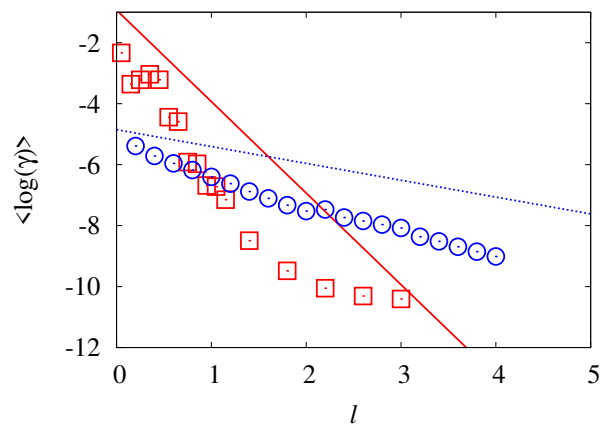}
  \caption{  (Color online)
    Ensemble averaged decay rates $\langle \log_{10} \gamma
    \rangle_s$ for $h=  1/5$ (square) and $1/20$ (circle). 
The theoretical lines $\gamma \sim e^{-2l/\xi}$ are also drawn. The undershoot in the plot of $h=1/5$ is an artifact of numerical fitting to the data from the limited time evolution, e.g. $n < 10^{7}$. When the width $l$ is large ($l>1$), the decay rates with a smaller Planck's constant ($h=1/20$) become larger than those with a larger Planck's constant. This is because the localization length $\xi$ increases as the Planck's constant decreases.
  }

  \label{FIG:NoScaled_l-Gamma}
\end{figure}
In the cases considered so far, noise was applied to the whole chaotic
sea, that is $L \approx d$. In this section, we see how the decay rate
$\gamma$ behaves and how the localization length $\xi$ depends on
$\gamma$ when noise is applied only to the limited chaotic region
which is distant from the torus region, that is $L>d$. In the
following we denote the width of the unperturbed chaotic region, that
is the region without noise, by $l = L- d$
as shown in Fig.~\ref{FIG:Schematic} schematically. 

For $l >0$, since dynamical localization remains in the unperturbed
chaotic region $d<p<L$, we can expect that dynamical tunneling is
suppressed partially by the remaining localization. On the other hand,
in the perturbed region $p>L$, we can expect that noise destroys
dynamical localization to restore the diffusion process for the same
reason as discussed above. The latter is again supposed to help the
restoration of the chaotic tunneling. In order to avoid complications,
we here restrict ourselves to the case where the noise intensity is
sufficiently large, which corresponds to 
the case with large $\epsilon$ examined in the previous section.

As is shown in Fig.~\ref{FIG:Gamma-L}, we found that $\gamma$ tends to
decrease  exponentially as $l$ increases. Furthermore, when $l$ is
larger than a certain value, $\gamma$ strongly fluctuates with respect
to $s$. An interpretation for the presence of large fluctuation is
that the resonance previously observed in the perturbative regime
(i.e., smaller $\epsilon$) with $l\approx0$ is recovered. It should also be noted that $\gamma$ is not monotonic with $l$. We suppose that this is also due to the recovery of the resonance, whose quasienergy gap is modulated by $l$.

In the following, we discuss a gross feature of the $\gamma$-dependence on $l$ through an ensemble average with respect to $s$ within a narrow interval $\Delta s = 0.05$.
Since the fluctuation of $\gamma$ is rather large, we take the average $\langle\ln\gamma\rangle_s$, instead of  $\langle\gamma\rangle_s$. As shown in Fig.~\ref{FIG:Gamma-L}(b), the average $\langle\ln\gamma\rangle_s$ decreases monotonically as a function of $l$. This is consistent with exponential dependence of $\gamma$ on $l$. On the other hand, the fluctuation of $\ln\gamma$ gradually becomes larger as $l$ increases. This is due to the presence of resonances as explained above.

The exponentially decaying behavior of $\gamma$ as a function of $l$ suggests the presence of a characteristic momentum scale.  It is simplest to hypothesize the following dependence of $\gamma$ on the localization length $\xi$ of the unperturbed chaotic region.  \begin{equation}
 \label{eq:AverageLnGamma}
 \langle \ln \gamma \rangle_s \sim \ln \gamma_0 -\alpha \frac{l}{\xi }
\end{equation}
where $\gamma_0$ and $\alpha$ are constants.

To examine the validity of this hypothesis, we first determine the localization length $\xi$ by fitting the slope of the probability distribution numerically. The value of $\alpha$ can be roughly estimated using the following perturbative argument. Let a wavepacket evolve from the torus without noise. If we wait long enough, the wavepacket becomes a superposition of the torus state and a chaotic state. The latter exhibits dynamical localization in the chaotic sea (see, Sec.~\ref{sec:CAT}). In this case, the envelope of the wave function in the momentum representation takes the exponential form $\simeq e^{-p/\xi}$ in the chaotic sea. Once we impose noise at $p > L$, tunneling penetration starts. We assume that the wave function for $p < L$ is almost unchanged in spite of the presence of noise. Then the rate of the tunneling would be proportional to $|e^{-l/\xi}|^2$. This implies $\alpha=2$. A test of our argument for $h=1/20$ is shown in Fig.~\ref{FIG:Gamma-L}(b). The result suggests $\alpha$ should be slightly larger than $2$.

In Fig.~\ref{FIG:scaled-gamma}, the validity of Eq.~\eqref{eq:AverageLnGamma} for various values of $h$ is examined. If $\alpha$ in Eq.~\eqref{eq:AverageLnGamma} is assumed to be independent of $h$, $\gamma$ depends on $h$ only through $\gamma_0$ and $\xi$. Hence we check how the scaled decay rate $\langle \ln (\gamma/\gamma_0) \rangle_s$ depends on the scaled width of the unperturbed chaotic region $l/\xi$. The results in Fig.~\ref{FIG:scaled-gamma} roughly validate our assumed scaling relation Eq.~\eqref{eq:AverageLnGamma} with $\alpha$ slightly less than 2. We thus conclude that the localization length is a crucial factor for determining $\gamma$. In other words, chaotic tunneling is shown to be strongly correlated with dynamical localization in such a way.

We discuss an implication of the scaling
relation~\eqref{eq:AverageLnGamma} for $h$-dependence on $\gamma$. On
one hand, $\gamma_0$ decreases as $h\to 0$. On the other hand, the
localization length $\xi$ increases as $h\to 0$ \cite{Shep}. Hence,
$h$ dependence of $\gamma \simeq \gamma_0 \exp(-\alpha l /\xi)$ for a
given value of $l$ is not monotonic, provided $l>0$. An example is
shown in Fig.~\ref{FIG:NoScaled_l-Gamma}, where $l$ vs. $\langle \ln
(\gamma) \rangle_s$ is plotted for two values of $h$. For small $l$,
$\langle \ln (\gamma) \rangle_s$ becomes smaller as $h$ becomes
smaller.
This is explained by $h$-dependence of $\gamma_0$. For large $l$, the
opposite occurs. Namely, the tunneling rate grows even though the
Planck's constant $h$ is reduced. 
Hence we discovered that the Planck's constant dependence of the
tunneling rate is not monotonic when there partially remains dynamical
localization 
in the chaotic sea.  This is quite a unique
feature and another piece of evidence for the strong correlation
between dynamical localization and chaotic tunneling.

\section{The case with smoothed border}
\label{sec:Diffraction}

\begin{figure}[tb]
  \centering
  \includegraphics[width=8.5cm, clip]
   {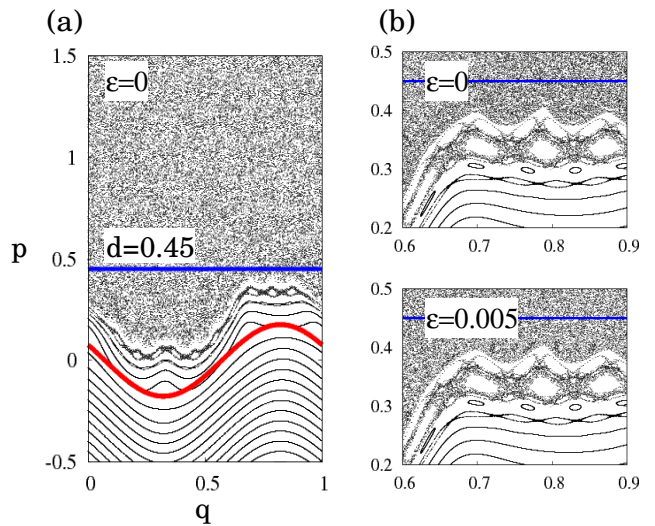}
  \caption{%
    (Color online)
Phase space portraits for small $\beta$ ($=5$), where the border
between torus and chaotic regions is blurred, and island chains appear
around the border. Other parameters are given as $d=0.45$ and $s=4$. A
full view and its magnification are shown in (a) and (b),
respectively. In (a), we depict a Lagrangian manifold corresponding to
the initial wave function $\PsiTorus{}(q; p_{\rm c})$ with $p_c = 0$ by
a thick curve. Also, a thick horizontal line at $p=d$ indicates the
discontinuity of the smoothed step function that appears in the
kinetic term $T_0(p)$. When $\beta$ is large, the line $p=d$ touches
the border (see Fig.~\ref{FIG:PhaseSpaces_s}). On the other hand, for
small $\beta$, the line $p=d$ and the boundary are detached \cite{EN6}. (b) Magnifications around the border without noise ($\epsilon=0.0$, $L=d+0.2$) and with noise ($\epsilon=0.005$, $L=d+0.2$). Note that there are almost no effects of noise around the border.
  }
  \label{FIG:DeformedPhaseSpace}
\end{figure}
\begin{figure}[tb]
  \centering
  \includegraphics[width=8.5cm, clip]
   {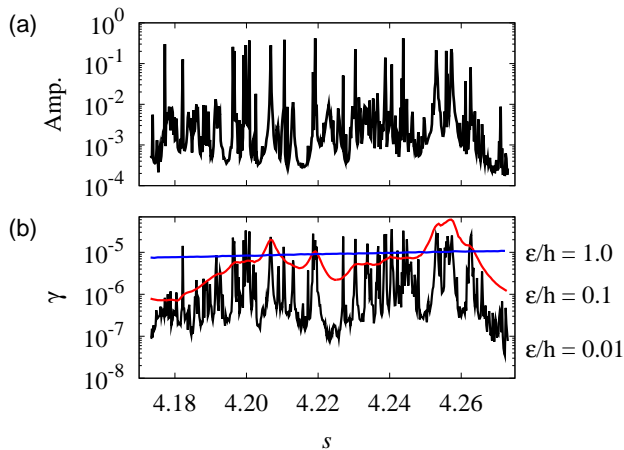}
  \caption{%
    (Color online)
Dependence of amplitude of tunneling oscillation and tunneling decay rate on parameter $s$ in the case of blurred border (a) amplitude of tunneling oscillation, (b) tunneling decay rate $\gamma$. The parameters of the underlying classical systems are the same as in Fig~\ref{FIG:DeformedPhaseSpace}. Here we take $h=1/20$. In (b), the strengths of $\epsilon$ are the same as in Fig.~\ref{FIG:Gamma_s_Epsilon}~(b).
  }
  \label{FIG:S-AMP-GAMMA_DeformedCase}
\end{figure}

Up to now, we have studied the system whose phase space is sharply separated between torus and chaos, and the degree of sharpness has been controlled by the parameter $\beta$. This is motivated, as stated in section~\ref{sec:phasespaceenginnering}, to remove complicated effects caused by the presence of complicated invariant structures in generic phase space. However, as a price we have to pay, the sharpness of the border between torus and chaotic regions may induce anomalous effects in quantum mechanics. 

What is important to realize a sharp border is to take the kinetic
term $T_0(p)$ with discontinuity (Eq.~\eqref{eq:defT0}). Because of
the nature of the $\theta_{\beta}(x)$ presented in
\eqref{eq:defTheta}, the second derivative $T''_0(p)$ is nearly
discontinuous. Especially, in the limit $\beta\to\infty$ $T'_0(p)$ is
strictly sharp and thus the free evolution gives rise to the so-called
Gibbs phenomenon in the Fourier transformation of the wave function in
the momentum representation. Namely, the wave function is strongly
diffracted \cite{Keller}. Furthermore, since the discontinuity of $T''_0(p)$ lies near the
border through which dynamical tunneling proceeds, diffraction can
have strong influence on the tunneling process. In our calculations,
we have used a large, yet finite $\beta$ (say, $100$ in
Fig.~\ref{FIG:PhaseSpaces_s}) , but the observation of Husimi
functions in the first several steps clearly reveals that there
remains a signature of diffraction even in case $\beta = 100$ \cite{ITS1}.
Therefore we here examine whether or not what has been claimed in the previous sections is valid even in small $\beta$ cases, where the signature of diffraction is expected to be negligible. 

In the case of large $\beta$ shown in previous sections, a torus state
and a localized state in the chaotic sea play a central role. As
$\beta$ becomes small, resonance islands  appear around the border and
the island states associated with them need to be taken into account
in addition to torus and chaotic states \cite{Frischat95,Ketzmerick00,BSU}. This may invoke two kinds of ``tunneling channels'' (1) a torus to chaotic sea; (2) a torus to an island, first, and subsequently the island to chaotic sea. 

Our scenario for the numerical study of small $\beta$ cases is almost the same as before: launch a wave function from an approximate torus state $\PsiTorus{}(q; p_{\rm c})$ and monitor the time evolution of $\Ptorus$. We add noise in a similar manner as before to destroy the quantum coherence producing dynamical localization. 

For $\beta =5$, as shown in Fig.~\ref{FIG:DeformedPhaseSpace}, there emerges nonlinear resonances in the torus region and small islands in the chaotic sea as  expected. It is important to note that small islands thus created are almost unchanged   even though noise is applied in the region $p > d$. An example of a phase portrait is presented in Fig.~\ref{FIG:DeformedPhaseSpace}~(b). This suggests that the changes of the small islands at different noise strengths do not have any influence on the tunneling process. 

In the absence of noise, whether the quasienergy resonance occurs or not, the wave function is almost bounded in the torus region and only a small fraction of tunneling tail penetrates the dynamical barrier. Thus, the behavior of $\Ptorus$ is similar as that shown in Figs.~\ref{FIG:TimeEvo-NonReso} and \ref{FIG:TimeEvo-Reso}. As is seen in Fig.~\ref{FIG:S-AMP-GAMMA_DeformedCase} (a), the amplitude of the tunneling oscillation is also sensitive to the variation in the parameter $s$. 

Note that the sensitivity with respect to $s$ is not reduced even with smaller $\beta$. The origin of the sensitivity should be attributed to the resonance among approximate stationary states, as explained in Section~\ref{sec:p2pi}. We comment on the contribution from the island states, which was not taken into account in the explanation in Section~\ref{sec:p2pi}. This is expected to be small, since the variation in $s$ influences only on the region far apart from the torus region,
and the quasienergies of island states do not depend on $s$. If a quasienergy resonance is sensitive with respect to $s$, it must involve a localized state in the chaotic sea. Hence we conclude such a resonance is the origin of the strong fluctuation.

As shown in Fig.~\ref{FIG:S-AMP-GAMMA_DeformedCase} (b), we have confirmed that 
chaotic tunneling is restored even for small $\beta$, and that the
behavior of the decay rate $\gamma$ also persists.
We note that $\gamma$ for small $\beta$ needs to be deliberately
determined from the time series of $\Ptorus$ because the deformation
of the tori near the border makes the preparation of an initial
torus state difficult \cite{EN7}.
In the present numerical experiment, we again take $\PsiTorus{}(q; p_{\rm c})$ as an initial state which is independent with $\beta$  (see Appendix~\ref{sec:QuantizeTorus}, with $p_{\rm c} = 0$.). The discrepancy of $\PsiTorus{}(q; p_{\rm c})$ from an approximate eigenstate on the torus produces
multiple characteristic time scales in $\Ptorus$, each of which is
associated  with an individual torus state. Here we estimate $\gamma$
from the most significant contribution in $\Ptorus$ \cite{EN7}. Evidence justifying such a procedure is shown in Fig.~\ref{FIG:S-AMP-GAMMA_DeformedCase}(b): the positions of oscillation peaks for the case without noise $\epsilon=0$ are in good agreement with those for the case with small $\epsilon$. With increase in $\epsilon$, we can see that sensitive dependence of $\gamma$ gradually disappears, and $\gamma$ becomes larger on average. All these are consistent with the results for large $\beta$. 

Here we show another reasoning that the deformation of islands with the changes of $s$ does not induce the sensitive fluctuations of the amplitude of the tunneling oscillations and $\gamma$ in Fig.~\ref{FIG:S-AMP-GAMMA_DeformedCase}. First, the islands scarcely change with noise, as is seen in Fig.~\ref{FIG:S-AMP-GAMMA_DeformedCase}.  This suggests that the influence of the $s$-dependence of the islands on the fluctuations is almost independent of the strength of noise. On the other hand, the fluctuation in $\gamma$ disappears with sufficiently strong noise. Hence we conclude that these fluctuations sensitive to $s$ are mainly due to the quasienergy resonance.

\section{Discussion and summary}
\label{sec:discussion}

In this paper, we have shown that there exists a very close relation between dynamical tunneling and dynamical localization. As mentioned in the introduction, chaos-assisted tunneling or chaotic tunneling is usually understood as a mechanism which enhances tunneling. It is certainly true that the transition between quasi-doublets, whose support in phase space is symmetric tori, acquires additional tunneling pathways with the assistance of chaotic states. Semiclassical arguments also provide a similar picture: exponentially many tunneling paths connect the torus region to the chaotic sea. 

On the other hand, what we have seen in the present paper is another
aspect of the role of chaos in the tunneling process. Since the
tunneling effect is a typical non-local wave phenomenon, it is natural
to expect that what is happening in chaotic regions must affect the
nature of the wave function in the integrable region and vice versa. It
is well known that even without stable invariant components,
wave-functions in the chaotic sea are localized in nontrivial ways,
e.g., scarring due to unstable periodic orbits \cite{Heller}, localization on cantori, and dynamical localization. One may thus ask the question how and to what extent such wave effects appearing in the chaotic region affect or are influenced by dynamical tunneling between torus and chaos.  

Here we focused particularly on the interplay between tunneling and
dynamical localization and gave evidence that
the non-local nature of quantum tunneling
becomes manifest if one of the quantum effects, that is
dynamical localization, is attenuated. If we add noise to the chaotic
region, which destroys the interference generating dynamical
localization, drastic enhancement of the tunneling rate is observed. 

The result may be interpreted in two different ways: on one hand, we may simply say that chaotic tunneling is strongly amplified under the influence of noise. On the other hand, we may regard the drastic enhancement of the tunneling rate as a result of manifestation of {\em potentially existing} chaotic tunneling. 

The former interpretation is a bit too naive, but most arguments
concerning chaotic tunneling so far have been based on this
picture. This implicitly assumes that chaotic tunneling is already
present even without noise and it occurs {\em irrespective of the
  status of surrounding chaotic seas}.
In this picture,
the tunneling process ends when the wave function goes through KAM
barriers, and 
the behavior
shown in Fig.~\ref{FIG:TimeEvo-NonReso} is viewed as manifestation of
chaotic tunneling.

In contrast, a semiclassical argument employing complex classical
orbits strongly supports and is consistent with the latter
interpretation. As mentioned in the introduction, complex orbits
contributing to the tunneling amplitude have an amphibious
character. They play the role of tunneling orbits when they move into
the integrable domain, but behave as almost real orbits after escaping
out of the integrable domain. There are exponentially many complex
orbits even in the integrable regime, and they flow into the chaotic
sea \cite{SI02}. Since these orbits are almost completely governed by
the real dynamics in the chaotic sea, the diffusive motion is
suppressed by dynamical localization as in the genuine real
dynamics. In this sense we can say that the tunneling decay is
inhibited due to dynamical localization in the chaotic sea, and
exponentially many potentially existing tunneling orbits appear if
noise destroys the coherence creating localization. If this
interpretation is true, it should be possible to explain the maximal
tunneling rate, which was observed in the strong noise limit (see
Fig.~\ref{FIG:Gamma_s_Epsilon}) and called a ''classical'' tunneling
rate, by evaluating the sum over complex orbits which flow into the
chaotic sea, although this evaluation is still not achievable for
several reasons \cite{SII09}. 

We further note that, for almost the same reasons, the origin of coherent oscillation is not clearly accounted for within semiclassical arguments. As presented in Fig.~\ref{FIG:TimeEvo-Reso}, the quasi-doublet state was formed between one torus state and one chaotic state when scanning system parameters. It is reasonable to suppose that some simpler form of coherence than destructive interferences generating dynamical localization could underlie complex orbits connecting torus and chaotic regions, but the mechanism invoking the observed strong resonances is not well understood. 

The most noteworthy consequence of the interplay between dynamical tunneling and dynamical localization is that the increase in Planck's constant does not necessarily lead to the growth of tunneling rate. As Planck's constant gets larger, the localization length decreases, which acts to strengthen the effect of suppression by dynamical localization. If the suppression mechanism dominates the change in the tunneling rate, a counter-intuitive change in the tunneling rate with Planck's constant may happen. We verified in Fig.~\ref{FIG:NoScaled_l-Gamma} that such a range indeed exists. 

We finally discuss the sharpness of the border between torus and
chaotic regions. The motivation for constructing the system so as to
have sharply or clearly divided phase space was to avoid other complex
problems of generic phase spaces and make the study of our tunneling
problem simpler. In particular, the presence of nonlinear
resonances leads to resonance assisted tunneling \cite{BSU}, an
alternative mechanism enhancing the tunneling rate. Conversely,
cantori or broken tori play the role of partial barriers which
suppress the transport \cite{GRR, BW, CP}. It seems possible to avoid
such problems when we study a system with sharply divided phase
space. This is certainly the case in classical mechanics, but as
mentioned in section \ref{sec:Diffraction} the sharpness of the
boundary, more precisely the discontinuity of kinetic or potential
functions, necessarily invokes diffraction. In section
\ref{sec:Diffraction}, we have verified that the claims given for the
sharply divided case do not nevertheless need to be corrected at least
in a qualitative level. However, we should recall that the underlying
mechanism is quite different between diffraction and tunneling,
although both concern classically forbidden processes. We need
alternative semiclassical techniques to obtain quantitative
results. In particular, the semiclassical evaluation employing complex
classical orbits is no longer valid if the system loses analyticity.
It is also frustrating that the recently proposed formula to predict a
tunneling rate assumes the extension of
  tori of fictitious integrable system
 to the surrounding chaotic sea, which
 requires that the border between torus and chaotic regions be as
 sharp as possible \cite{BKLS}.  This alternatively means that
 the validity of this formula is maximally guaranteed in the strong
 diffraction limit. A quantitative evaluation of tunneling rate is not
 an easy task if we define dynamical tunneling as a process far from
 the diffraction limit. This issue will be dealt with in the
 forthcoming publication \cite{ITS1}.

Throughout this paper, noise is implemented as a stochastic process. It is easy to hypothesize that analogous phenomena will occur when we employ chaotic dynamics to realize noise. Thus we close this paper by raising the following question: is the spontaneous recovery of chaotic tunneling generic in multiple-dimensional systems with mixed phase space?

\appendix
\section{Quantization of integrable  tori} \label{sec:QuantizeTorus}
We show how we approximately construct a torus state, 
which is the initial state of our numerical experiments 
(see, Sec.~\ref{sec:CAT}).
In the torus region, the classical dynamics of
the system presented in Sec.~\ref{sec:referenceSystem}
is described by
\begin{align}
p_{n+1} &= p_n + k \sin (2\pi q_n),\\
q_{n+1} &= q_n + \omega.
\end{align}
We may assume that 
the momentum of an invariant torus is specified by a function of $q$
\begin{equation}
 \label{Pq_def}
 p=P(q).
\end{equation}
Namely, $(q, P(q))$ is supposed to be on a torus. After an
iteration of the integrable map, the system arrives at
$(q + \omega, P(q) + k\sin(2\pi q))$, which must agree with
$(q + \omega, P(q + \omega))$ by the assumption~\eqref{Pq_def}.
Hence $P(q)$ must satisfy the following equation
\begin{align}
 \label{eq:FEQforP}
 P(q) + k\sin(2\pi q) = P(q+\omega)
 .
\end{align}
To solve it, we expand $P(q)$ as
\begin{equation}
 \label{Pq}
 P(q) = \sum _{m=-\infty}^{\infty} a_m e^{2\pi i m q}.
\end{equation}
The functional equation~\eqref{eq:FEQforP} is cast into an algebraic
equation of $a_m$
\begin{align}
 \sum _{m} a_m e^{2\pi i m q} 
 + \frac{k}{2i} e^{2\pi i q} 
 - \frac{k}{2i} e^{-2\pi i q}
 =
 \sum _{m} a_m e^{2\pi i m \omega} e^{2\pi i m q}.
\end{align}
Hence we have 
\begin{equation}
P(q) = - \frac{k\cos (2\pi(q-\frac{\omega}{2}) )}{2\sin (\pi \omega)}
+ p_{\rm c}
,
\end{equation}
where an arbitrary constant $p_{\rm c}$ specifies a torus.
We remark that the solution of Eq.~\eqref{eq:FEQforP} exists only 
when $\omega$ is not any integer.

From the Einstein-Brillouin-Keller quantization, we obtain the quantized tori
\begin{align}
&
\PsiTorus{}(q; p_{\rm c})
\notag \\
&= A \exp \left[ \frac{2\pi i}{h}
  \int^{q}_{0} P(q') \, dq'
\right] \notag \\
&= A \exp \left[ \frac{2\pi i}{h}
  \left(
    \frac{-k}{4\pi\sin (\pi \omega)}
    \sin \left(2\pi (q-\frac{\omega}{2})\right) +
    p_{\rm c}q
  \right)
\right], \label{q_rep_torus}
\end{align}
where $A$ is a normalization constant.
Because $\PsiTorus{}(q; p_{\rm c})$ must be single-valued,
$p_{\rm c}$ must be quantized.


\begin{thebibliography}{24}
\bibitem{EN1}
  More precisely, $p_{\rm cutoff}$ is kept large enough  so as to
  preserve the dynamical localization that occurs in the vicinity of
  the torus region when the noise term is absent. As shown in
  Section~\ref{sec:expDecay}, we have non-zero decay rate from torus
  to chaotic region  under the presence of noise. Hence, we choose
  $p_{\rm cutoff}$ so as to obtain the converged decay rate.
\bibitem{EN2}
Precisely speaking, $\Ptorus$ also contains the probability at the
chaotic sea near the torus. However the contribution from the chaotic
sea is negligible on average, since probability in that region escapes
quickly.
\bibitem{EN4}
  The discontinuity of the lines of quasienergy
  is due to artifact of this condition.
\bibitem{EN5}
  As the position of the absorbing boundary is set to be
  $p_{\rm cutoff} > L$,  the probability amplitude obeys a diffusion process
  in the region  of $p_{\rm cutoff}> p > L $ and it escapes through the
  absorbing boundary.
\bibitem{EN6}
 The reason for the choice of $d$ in case of small $\beta$ is as
 follows. Since making $\beta$ small makes the border move
 ``downward'' (i.e., toward smaller $p$), the Lagrangian manifold of
 the initial state may cross the border. In order to avoid this, we
 have to move the border upward or the Lagrangian manifold
 downward. The latter is infeasible because it is impossible to move
 continuously the Lagrangian manifold due to the quantization
 condition of $\PsiTorus{}(q)$. Hence we move the border upward by
 increasing $d$.
\bibitem{EN7}
  The decay rate $\gamma$ is determined from $\Ptorus$ that
  are not too small ($\Ptorus>0.1$) and $n\le 10^5$,
  by least square fitting.
\end{thebibliography}
\end{document}